\documentclass[]{elsart}

\usepackage{graphicx}

\def\bfi{\begin{figure}}
\def\efi{\end{figure}}
\def\bc{\begin{center}}
\def\ec{\end{center}}
\def\be{\begin{equation}}
\def\ee{\end{equation}}
\def\bitz{\begin{itemize}}
\def\eitz{\end{itemize}}
\def\benum{\begin{enumerate}}
\def\eenum{\end{enumerate}}

\begin{document}
\begin{frontmatter}

\title{Cathode Position Response of Large-Area Photomultipliers Under a Magnetic Field}

\author{T. Koblesky},
\author{J. Roloff},
\author{C. Polly},
\author{and J.C. Peng}

\address{Department of Physics, University of Illinois at Urbana-Champaign,
Urbana, Illinois 61801, USA}

\begin{abstract}
With the increasingly common use of large area PMTs (photomultiplier tubes) 
for nuclear and particle physics experiments, information on the 
position dependent 
magnetic field effects across the PMT's photocathode is important to 
effectively analyze and understand data collected from PMTs. Using an 
automated two-dimensional scanner, we have measured the PMT response 
to an external magnetic field as a function of the cathode position 
impinged by a collimated light source. This study shows a clear 
dependence on the cathode position in both collection efficiency and 
gain of the PMT in the presence of a magnetic field.  In particular, 
a pronounced valley in the collection efficiency is observed for 
certain locations of the cathode when the magnetic field is transverse 
to the PMT axis. The pattern of the position dependence is presented for 
several different magnitudes and orientations of the magnetic field 
relative to the PMT.
\end{abstract}

\end{frontmatter}

\section{Introduction}
\label{intro}

Large diameter hemispheric PMTs have been used extensively in many recent
neutrino experiments including Superkamiokande~\cite{superk}, 
Kamland~\cite{kamland}, SNO~\cite{sno}, MiniBoone~\cite{miniboone},
Icecube~\cite{icecube}, and Chooz~\cite{chooz}. Ongoing experiments
at Double-Chooz~\cite{dchooz}, DayaBay~\cite{dayabay}, and Reno~\cite{reno},
all aiming at precision measurements of the yet unknown neutrino mixing
angle $\theta_{13}$ using neutrinos from nuclear reactors, also
utilize 8-inch or 10-inch diameter hemispheric PMTs for neutrino
detection. In these precision neutrino oscillation experiments, the value
of the $\theta_{13}$ mixing angle will be extracted from a precise
comparison of the number of neutrino events measured at near versus
far sites. A good understanding of the performance of the PMTs is
crucial for minimizing the systematic uncertainties of these experiments.

In this article, we report results from a measurement of the response of 
an 8-inch diameter hemispheric PMT in the presence of a magnetic field.
The relatively large distance between the photo-cathode and the first
dynode makes a large diameter hemispheric PMT more prone to the
influence of ambient magnetic field, which would alter the trajectories of
photoelectrons emitted from the photo-cathode. As a result, the 
PMT efficiency, as well as the PMT gain, could be significantly 
affected by the presence of the magnetic field.

Several studies have already been reported in the literature on the
overall response of the PMT to magnetic field when light is shined
uniformly over the surface of the 
photo-cathode~\cite{icecube,miniboonepmt,dcpmt}. However, the 
dependence of the PMT response as a function of the cathode position 
in the presence of a magnetic field has not yet been reported.

The goal of this study is to obtain information on the 
cathode position dependence of the PMT response as a function of
the magnitude and orientation of the external magnetic field. This 
information could be used in a detailed modeling of the PMT response 
to magnetic field for analyzing the data, and could also help optimizing 
the design of future experiments utilizing large diameter hemispheric PMTs.

\section{Experimental Setup}
\label{expsetup}

\bfi
\includegraphics[height=9.cm]{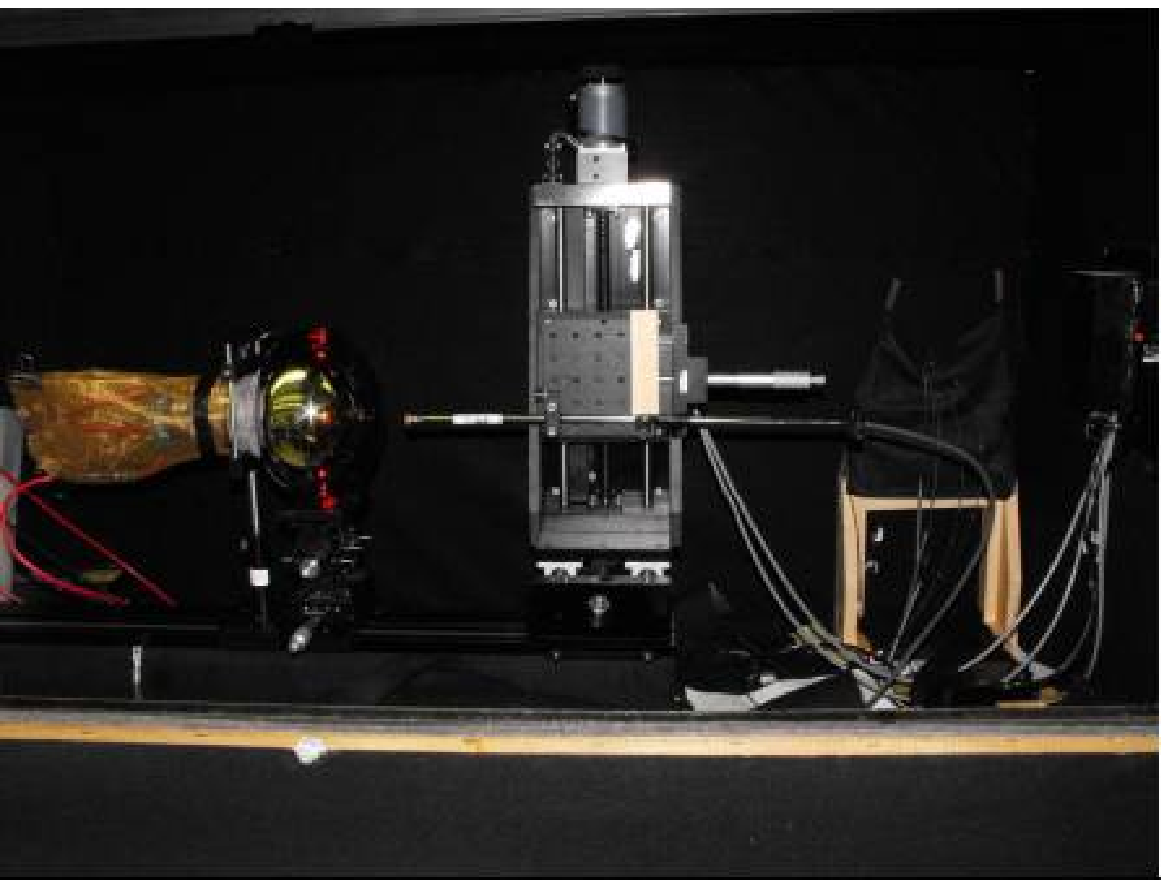}
\includegraphics[height=11.cm]{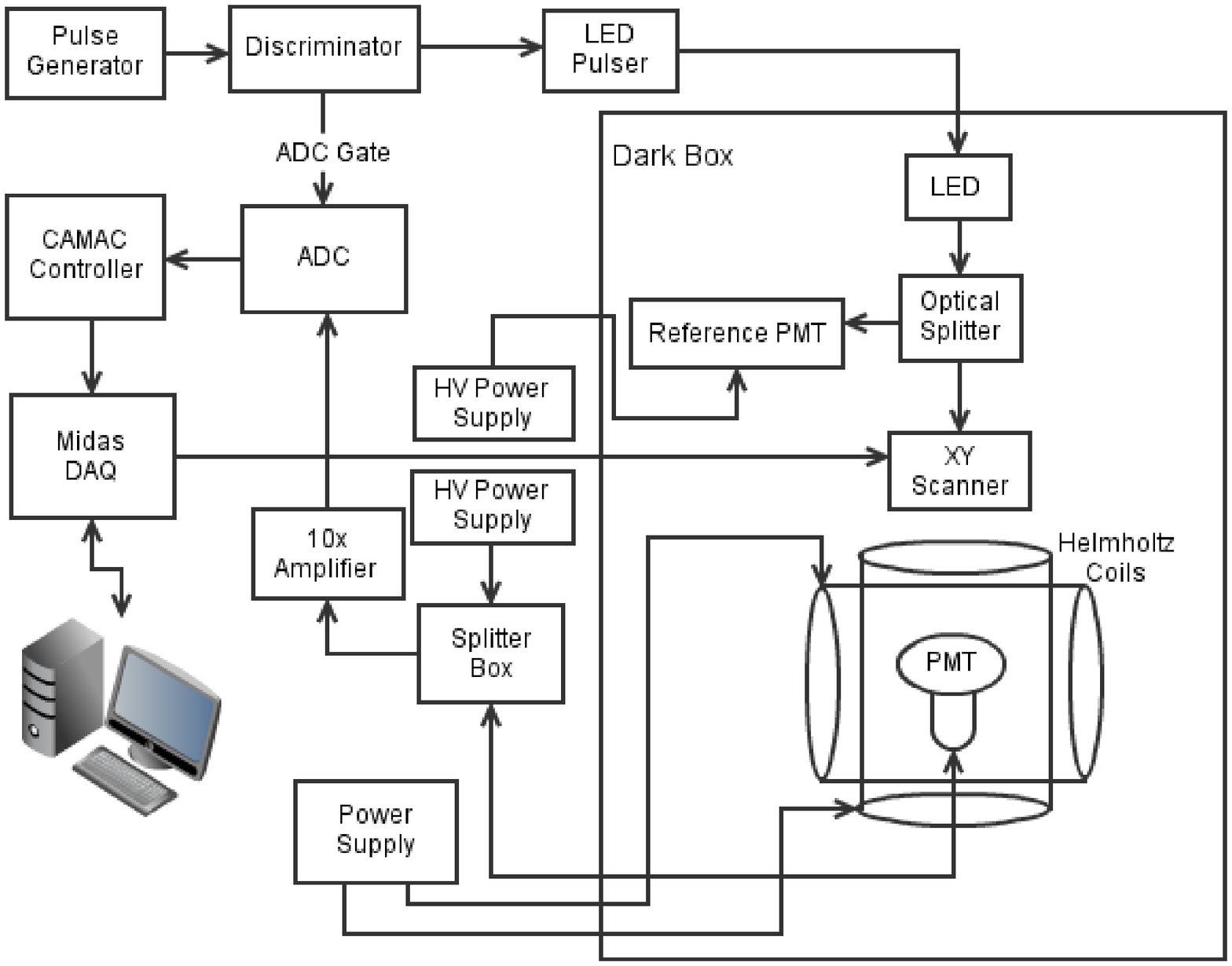}
\caption{a) A photo showing the PMT and the scanner in the dark 
box. b) The schematics of the setup of the experiment.}
\label{setup}
\efi

\bfi
\includegraphics[height=7.cm]{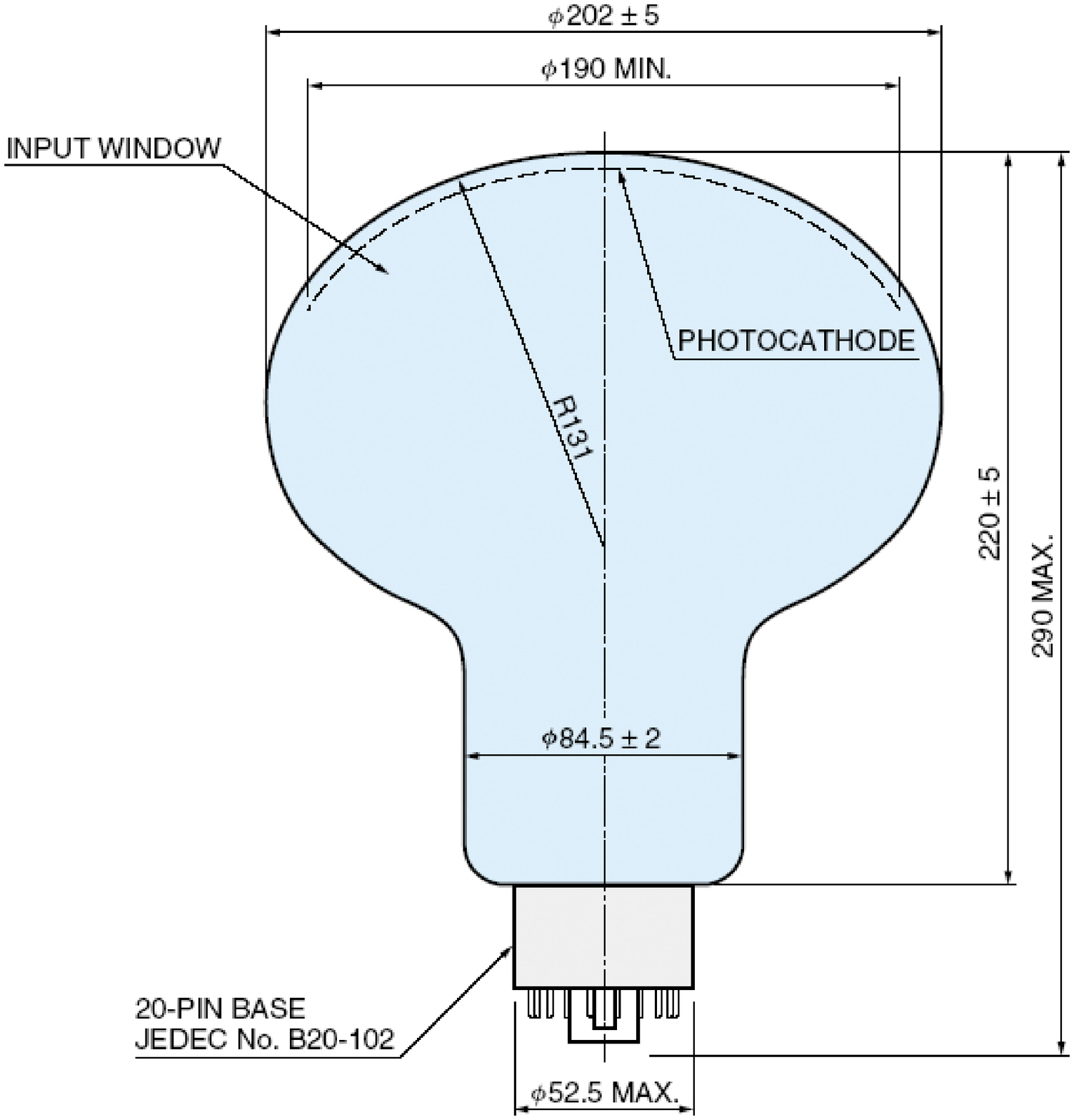}
\includegraphics[height=7.cm]{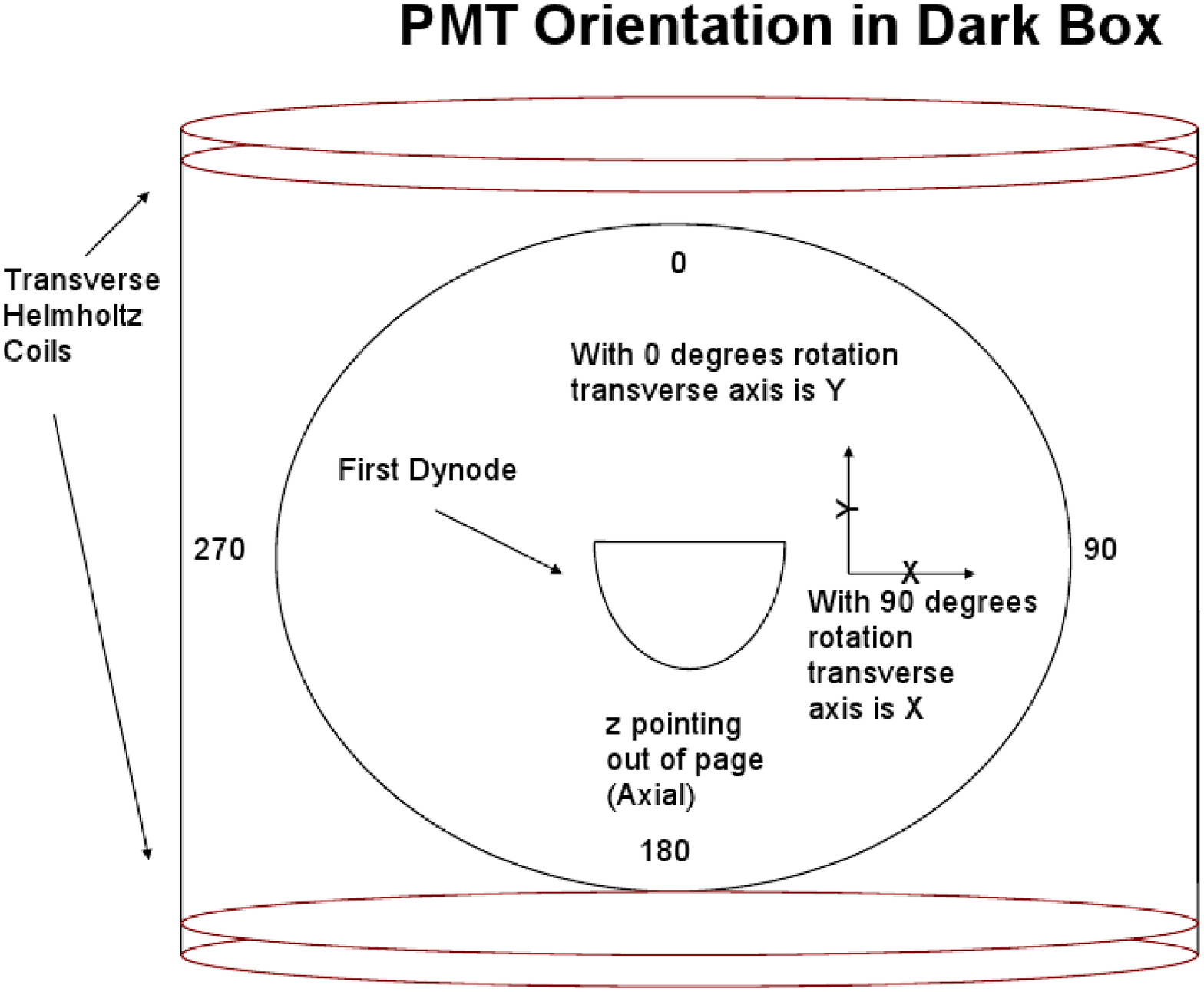}
\caption{a) Side-view of the Hamamatsu R5912 PMT. b) End-view of the PMT.
The x-y plane is also indicated}
\label{pmt}
\efi

Figure \ref{setup} shows the photo and schematics of the 
experimental setup. The PMT used in this study is an 8-inch Hamamastu 
R5912 \cite{r5912}. Similar models of PMTs have been used in several 
neutrino experiments including LSND, MiniBoone, and SNO, and will also 
be used in DayaBay.  For this study, a positive high voltage was applied 
to the PMT's anode. The PMT signal comes out of anode via the high 
voltage cable. In order to extract the signal, a splitter box was used 
to separate the high frequency components from the dc high voltage. 
The signal was fed into an ADC (analog-to-digital converter) to measure 
the integrated charge.

An LED, with a wavelength of 465 nm, chosen for its high quantum 
efficiency for this PMT, was pulsed with a frequency of 500 Hz. Although 
the source is fairly stable, the LED's light amplitude fluctuates slightly 
over the duration of a scan of the PMT. To correct for these fluctuations, 
three optical fibers diverted some of the LED light to a 2 inch diameter 
reference PMT shielded from magnetic fields in $\mu$-metal, while the 
remaining four optical fibers were directed towards the 8-inch PMT. The 
intensity of the LED can be monitored by the response of the reference PMT. 
The variation throughout the scan was minimal; in a 12 hour scan, the 
reference PMT recorded only a ~10$^{-3}$ variation of LED intensity. 
This small amount of fluctuation was removed in the analysis by dividing 
PMT's response by the reference PMT's response.  

In order to aim the fiber at various positions of the photo-cathode, 
an automated 2D scanner was used. This scanner consisted of a platform 
which could be moved to various (x,y) positions within range by 
stepping motors in two orthogonal directions. The x-y plane is perpendicular 
to the axis of the PMT, which is along the z-axis. The scanner has a range 
of 20 cm in both directions perpendicular to the z-axis with a minimum 
step size of 1 $\mu$m~\cite{scanner}. The PMT and optical fibers were 
oriented as shown in Figure \ref{setup}, and the fibers were positioned 
on the scanner so that it would move up, down, left, and right to aim at 
various points on the PMT photo-cathode. The fibers were positioned ~2 
inches away from the photo-cathode, with an average circular spot size 
of 2 mm.

Two pairs of 45-cm diameter Helmholtz coils were positioned around the 
PMT in order to cancel the earth's magnetic field or to induce magnetic 
fields of a given magnitude and direction. The first pair of Helmholtz 
coils was positioned along the z-axis, and the second along the y-axis. 
The two pairs of Helmholtz coils served also to cancel the earth's field 
of $-0.47$ gauss in the y-direction and 0.2 gauss in the z-direction.  
The earth's magnetic field along the x-axis was negligible. By inducing 
magnetic fields with these Helmholtz coils we were able to create magnetic 
fields within a range of magnitude in any direction in the y-z plane.

A Lecroy 2249W ADC was used to digitize the charges collected from the 
R5912 and the reference PMT.  The LED pulser triggered a discriminator 
that provided an ADC gate width of 55 ns. This width suitably contains 
the PMT signal which had a width on the order of 30 ns. The Midas DAQ 
system which interfaced with the CAMAC SCSI Bus Crate Controller(73A) 
was used to record data from the ADC. 

\section{Results of Measurements}
\label{measurement}

The effects of magnetic fields on the PMT collection efficiency and gain 
were measured at different orientations of the magnetic field with 
respect to the PMT.  The collection efficiency of a PMT refers to the 
probability of electrons emitted by the photo-cathode reaching the first 
dynode. The region between the photo-cathode and the first dynode is 
most susceptible to magnetic field effects. As shown in Figure \ref{pmt}b, 
the first dynode is not azimuthally symmetric. Hence the collection
efficiency is in general not azimuthally symmetric with respect to the
cathod position from which photo-electrons are emitted. This asymmetry
can be further enhanced in the presence of magnetic field. 

\bfi
\includegraphics[height=10.cm,width=15.cm]{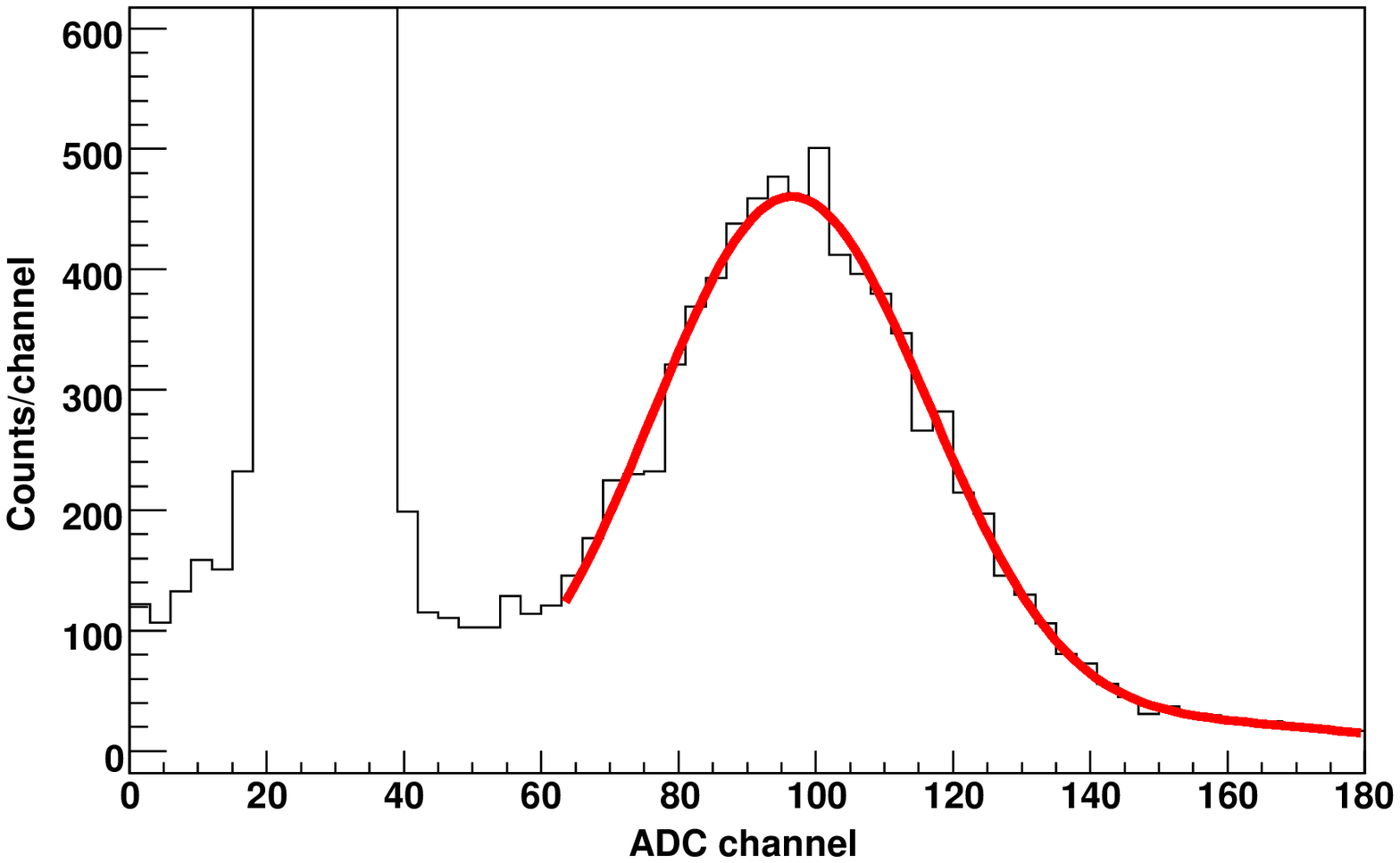}
\caption{A typical ADC spectrum showing the SPE and the pedestal peak.}
\label{spe}
\efi

To determine the collection efficiency and gain, we adjusted the LED intensity 
so that the single-photoelectron (SPE) peak is the dominant feature in the 
ADC spectrum. A typical ADC spectrum is shown in Figure \ref{spe}.  
The location of the SPE peak is fitted with a Gaussian distribution.  
After subtraction the pedestal channel from the SPE channel, the PMT gain 
is calculated taking into account the ADC calibration (0.25 pC/ch) and the 
effect of the terminator in the PMT base circuit. The high voltage was set 
at a PMT gain of $\sim 10^7$.

ADC spectra, similar to the one shown in Figure \ref{spe}, were also 
analyzed to extract information on the PMT collection efficiency.  
For a fixed LED intensity, the PMT detection efficiency is proportional 
to the mean value of photoelectrons, $\mu$, measured by the PMT.  
Using the Poisson distribution, \emph{P(k)}=$\mu^ke^{-\mu}/k!$, to 
describe the probability of detecting \emph{k} photoelectrons, one can 
readily calculate $\mu$ from \emph{P(0)}=$e^{-\mu}$.  The probability of 
detecting no photoelectron, \emph{P(0)} is determined from the number 
of pedestal events divided by the total number of events.

\bfi
\includegraphics[height=7.cm,width=7.cm]{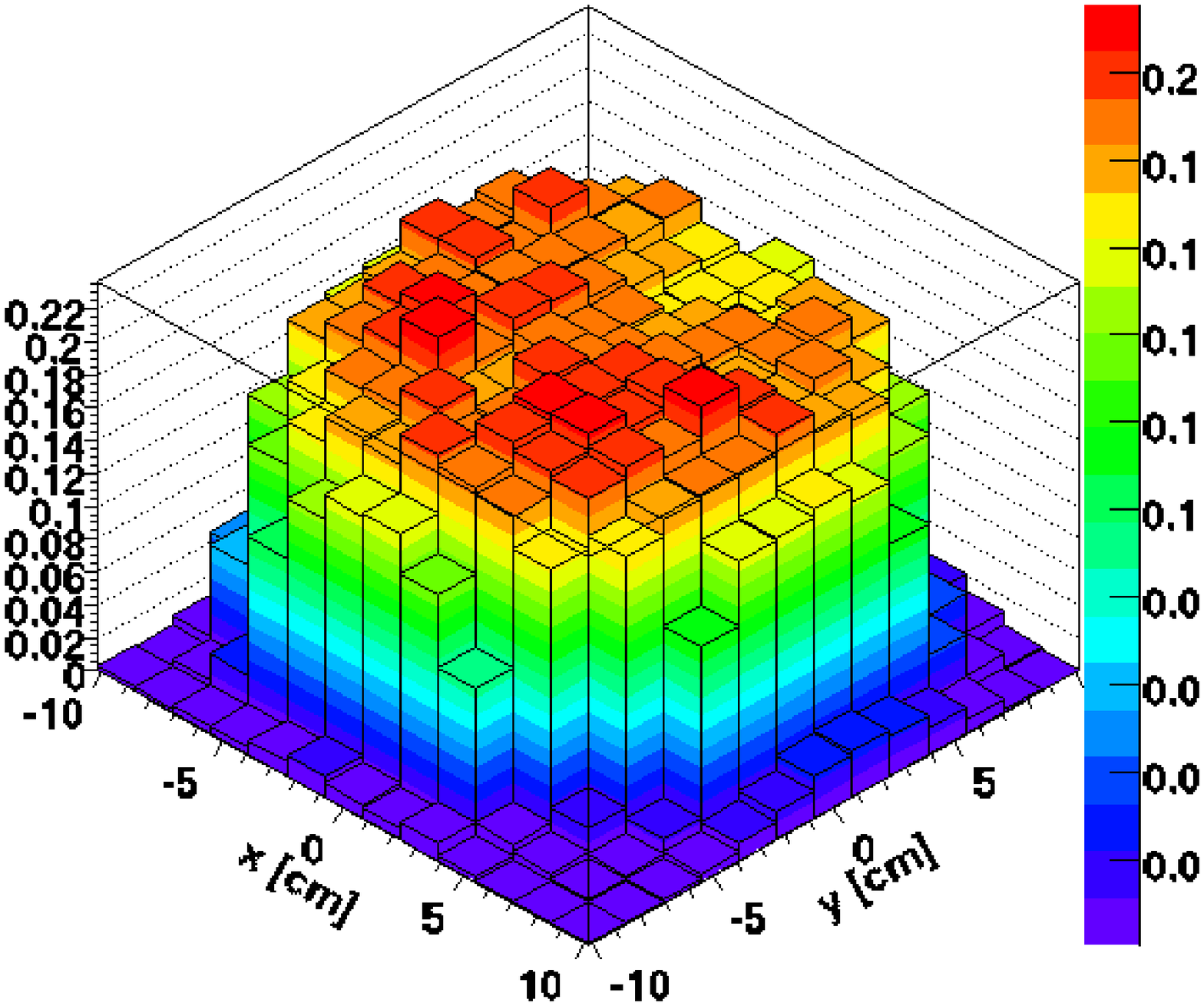}
\includegraphics[height=7.cm,width=7.cm]{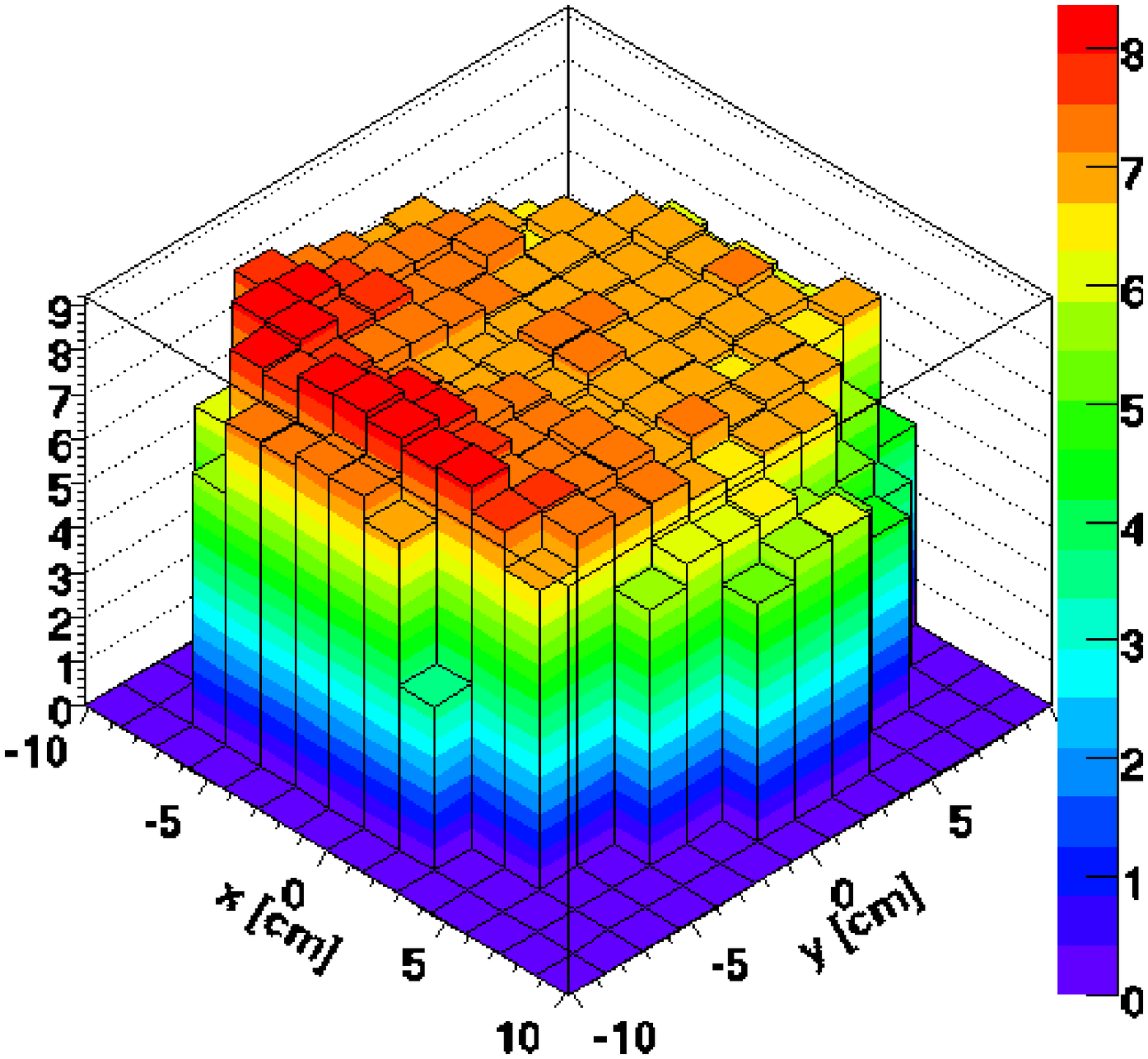}
\caption{a) Mean value of the photoelectron, $\mu$ and b) Gain of the 
PMT in units of log, of the PMT in zero magnetic field as a function 
of photo-cathode position.}
\label{zeroField}
\efi

To measure the PMT response to magnetic fields, we first measured its 
performance in zero magnetic field.  The earth's field (-0.47 gauss in the 
y-direction and 0.2 gauss in the z-direction) was cancelled by the two sets 
of Helmholtz coils producing opposite magnetic fields. 
Figure \ref{zeroField}a shows 
that $\mu$ is rather uniform across the cathode surface except near the 
edge, where reflection of LED light from the glass surface due to a larger 
incident angle led to lower collection efficiency. 
Figure \ref{zeroField}b shows that the PMT gain is largely independent
of the cathode position at zero magnetic field. Note that 
Fig. \ref{zeroField} is obtained by analyzing a large number of ADC 
spectra collected for various grid positions of the scanner.

\bfi
\includegraphics[height=7.cm,width=7.cm]{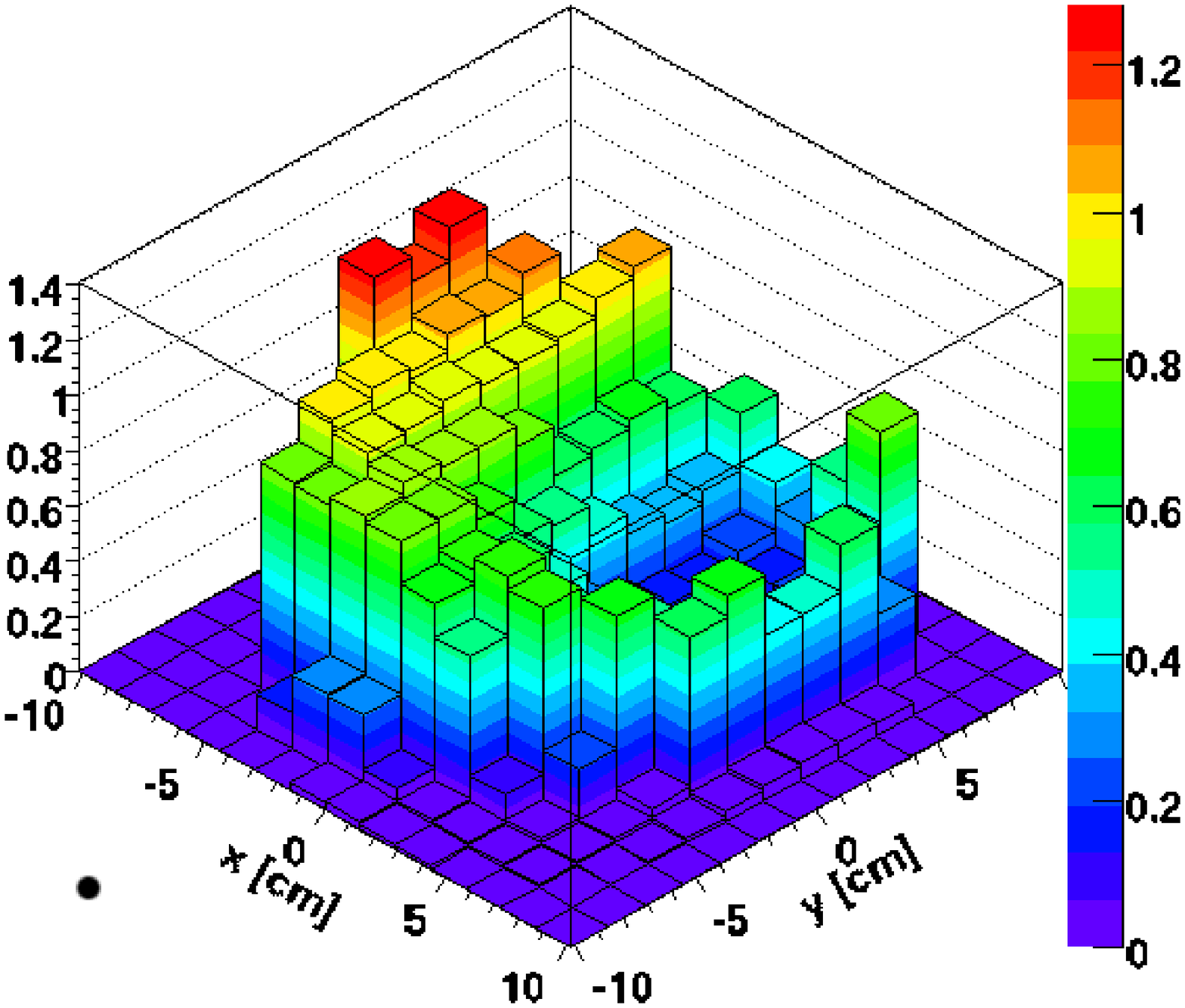}
\includegraphics[height=7.cm,width=7.cm]{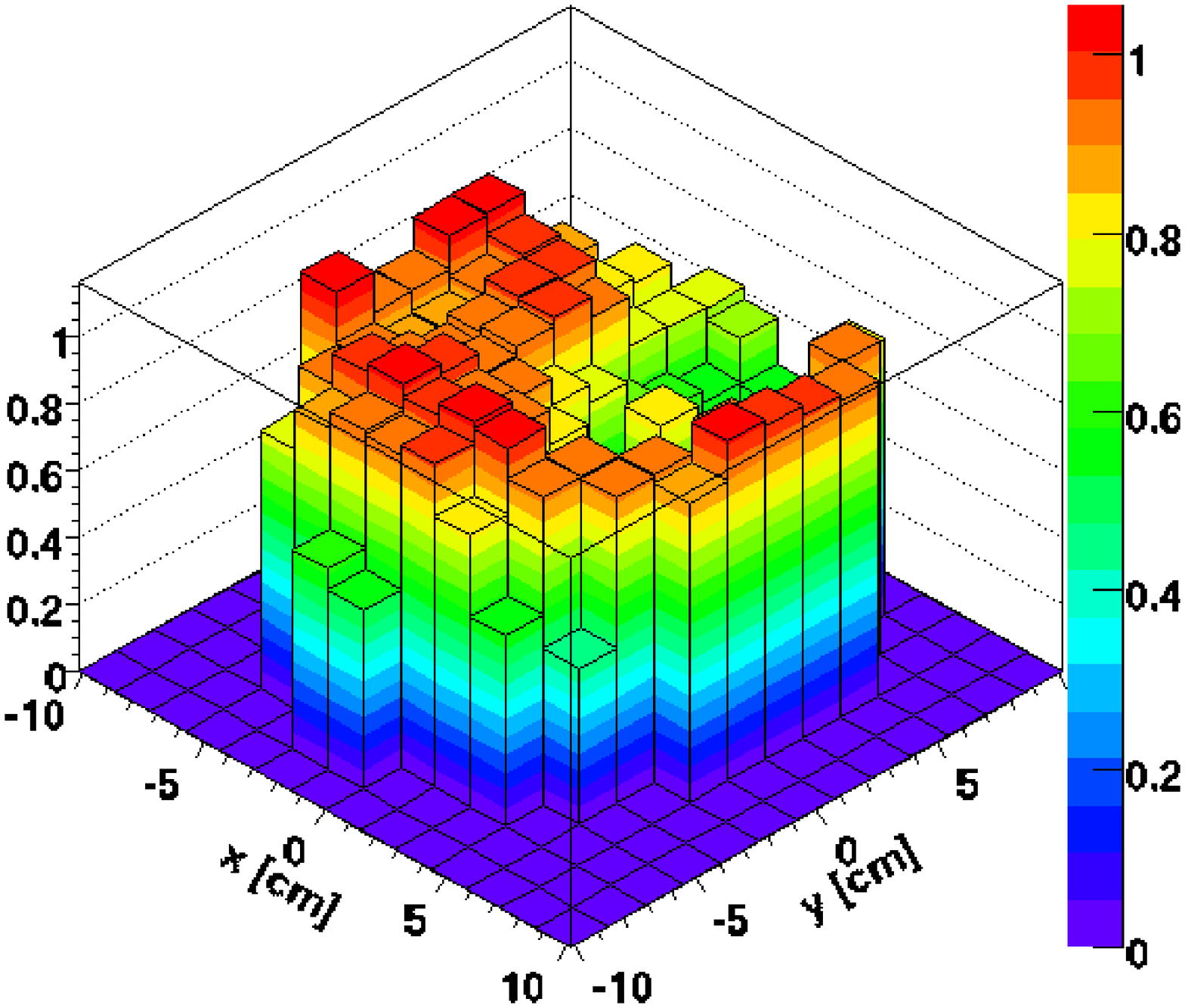}
\caption{Plots of a) The collection efficiency of the PMT b) The gain 
of the PMT with a $-0.5$ gauss magnetic field in the y-direction.}
\label{rot90_-.5GT}
\efi

The response of the PMT to a net external magnetic field has been measured 
with the magnetic field pointing perpendicular or parallel to the axis of 
the PMT.  Figure \ref{rot90_-.5GT} shows the PMT's relative collection 
efficiency $\epsilon$=$\mu/\mu_0$, when the net magnetic field is -0.5 
gauss in the \emph{y}-direction.  The relative collection efficiency is 
the rate of $\mu$ measured at a given magnetic field over the 
corresponding $\mu_0$ measured at zero magnetic field.  
Figure \ref{rot90_-.5GT}a shows that the presence of a magnetic 
field significantly modifies the PMT collection efficiency, and a striking 
position dependence is observed for $\epsilon$.  A pronounced valley 
in $\epsilon$ centered around \emph{x}=4 cm and \emph{y}=0 cm is observed.

Figure \ref{rot90_-.5GT}b shows the PMT relative gain, \emph{g}=\emph{G/$G_0$} 
as a function of the cathode position. \emph{G} is the PMT gain measured 
at a non-zero magnetic field, and \emph{$G_0$} is at zero magnetic field.  
As shown in Fig. \ref{rot90_-.5GT}, the effect of the magnetic field 
on PMT gain is much smaller than on the PMT collection efficiency.  
Nevertheless, \emph{g} also depends on the cathode position and has a less 
pronounced valley at the same region as that of $\epsilon$.  The observed 
cathode position dependence of the PMT gain implies that the mean deviation 
of the PMT gain would be larger in the presence of a magnetic field.

\bfi
\includegraphics[height=7.cm,width=7.cm]{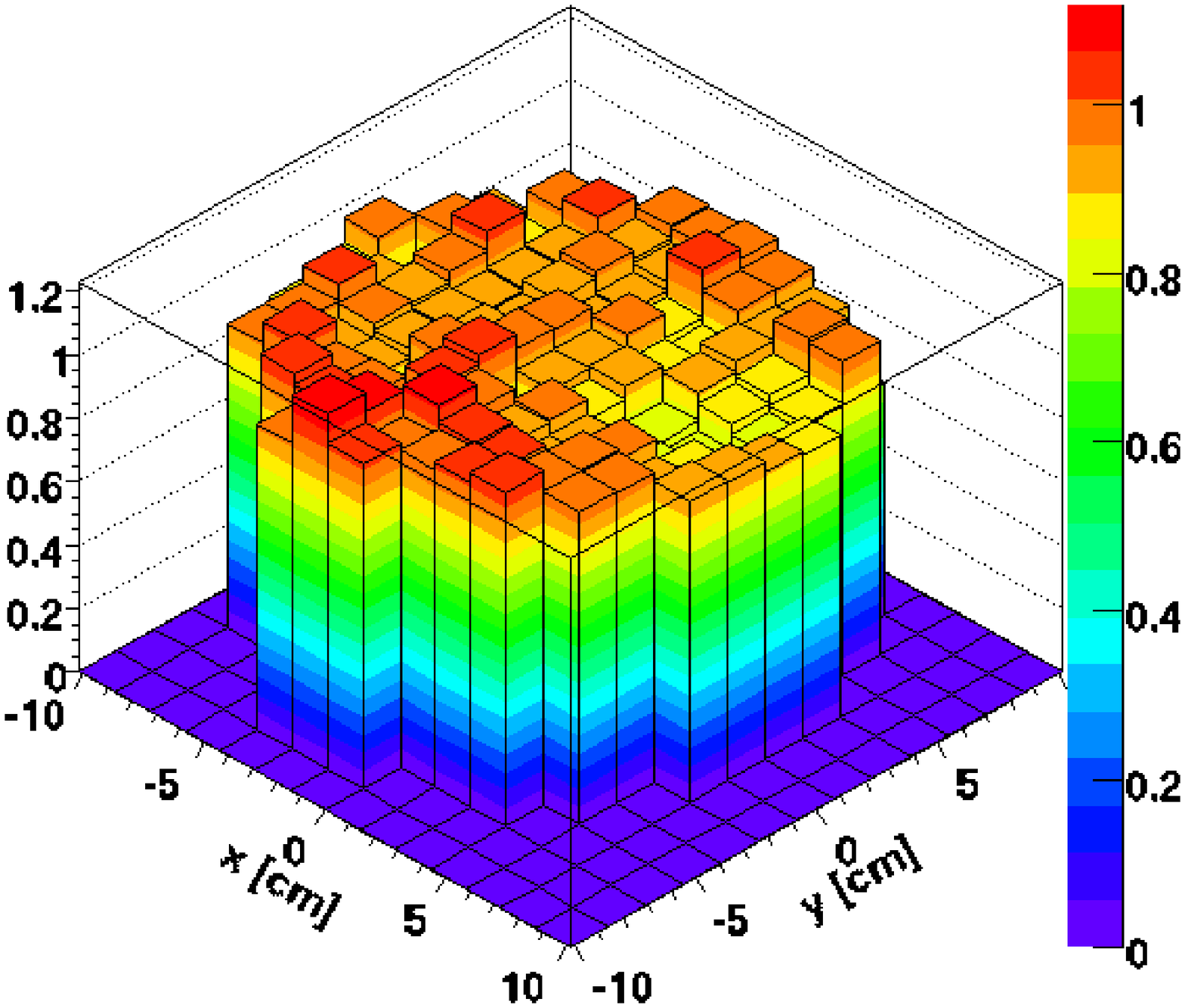}
\includegraphics[height=7.cm,width=7.cm]{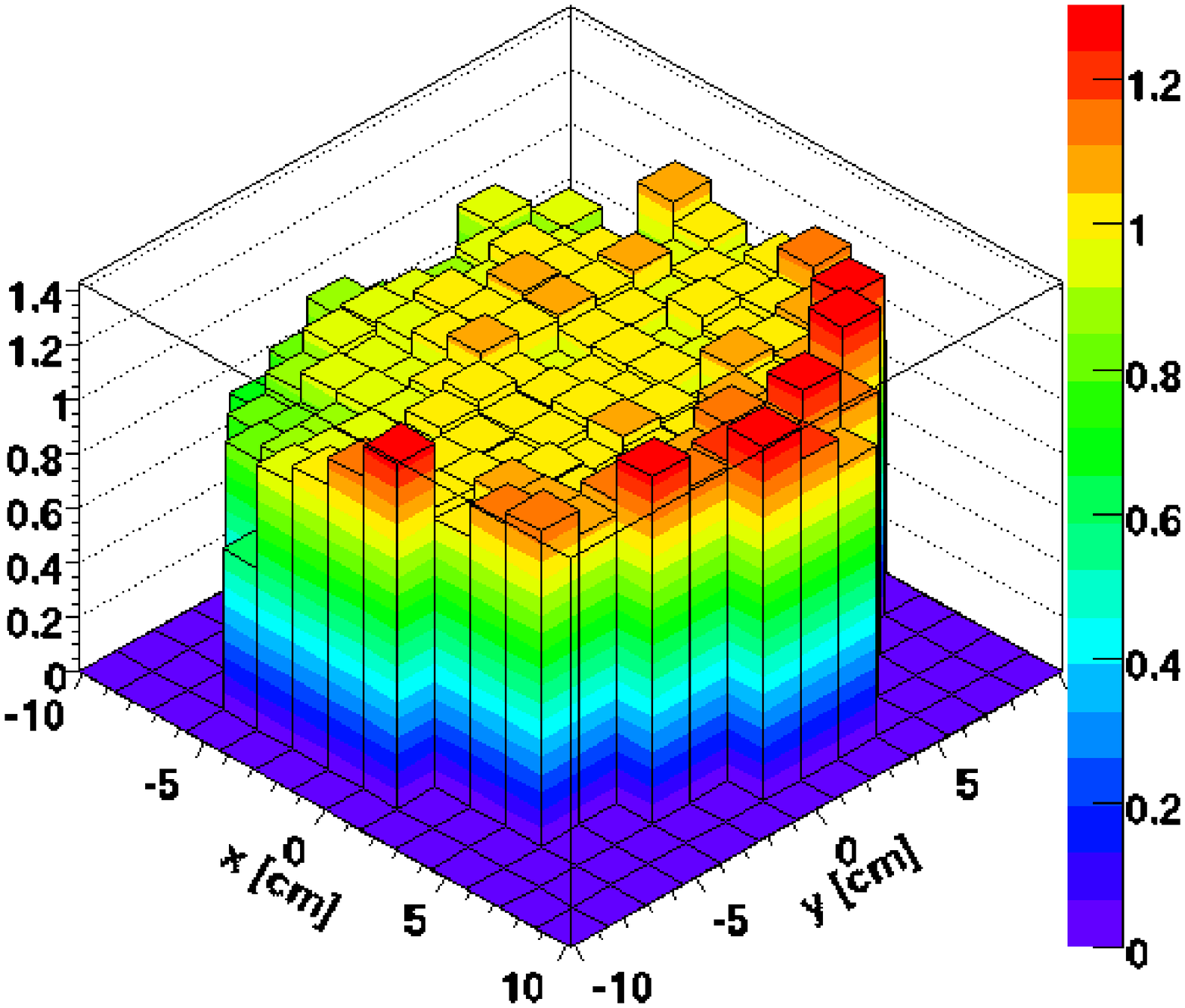}
\caption{Plots of a) The relative collection efficiency of the PMT b) The 
relative gain of the PMT for several points on the photocathode with 
a $-0.5$ gauss magnetic filed in the y-direction while the PMT is rotated
by 90$^\circ$.}
\label{rot90_all}
\efi

Figure~\ref{rot90_all} shows the PMT response when the PMT is rotated by
90$^\circ$ around its axis.
There is a notable difference between the response of the PMT in 
Figure \ref{rot90_-.5GT} and Figure \ref{rot90_all} when considering 
that the only difference between the plots is that the magnetic 
field has been rotated by 90$^{\circ}$. The pronounced hole that appears 
in the collection efficiency $\epsilon$ of Figure \ref{rot90_-.5GT}a 
disappears when the transverse magnetic field is rotated by 90$^{\circ}$ 
as seen in Figure \ref{rot90_all}a.

\bfi
\includegraphics[height=7.cm,width=7.cm]{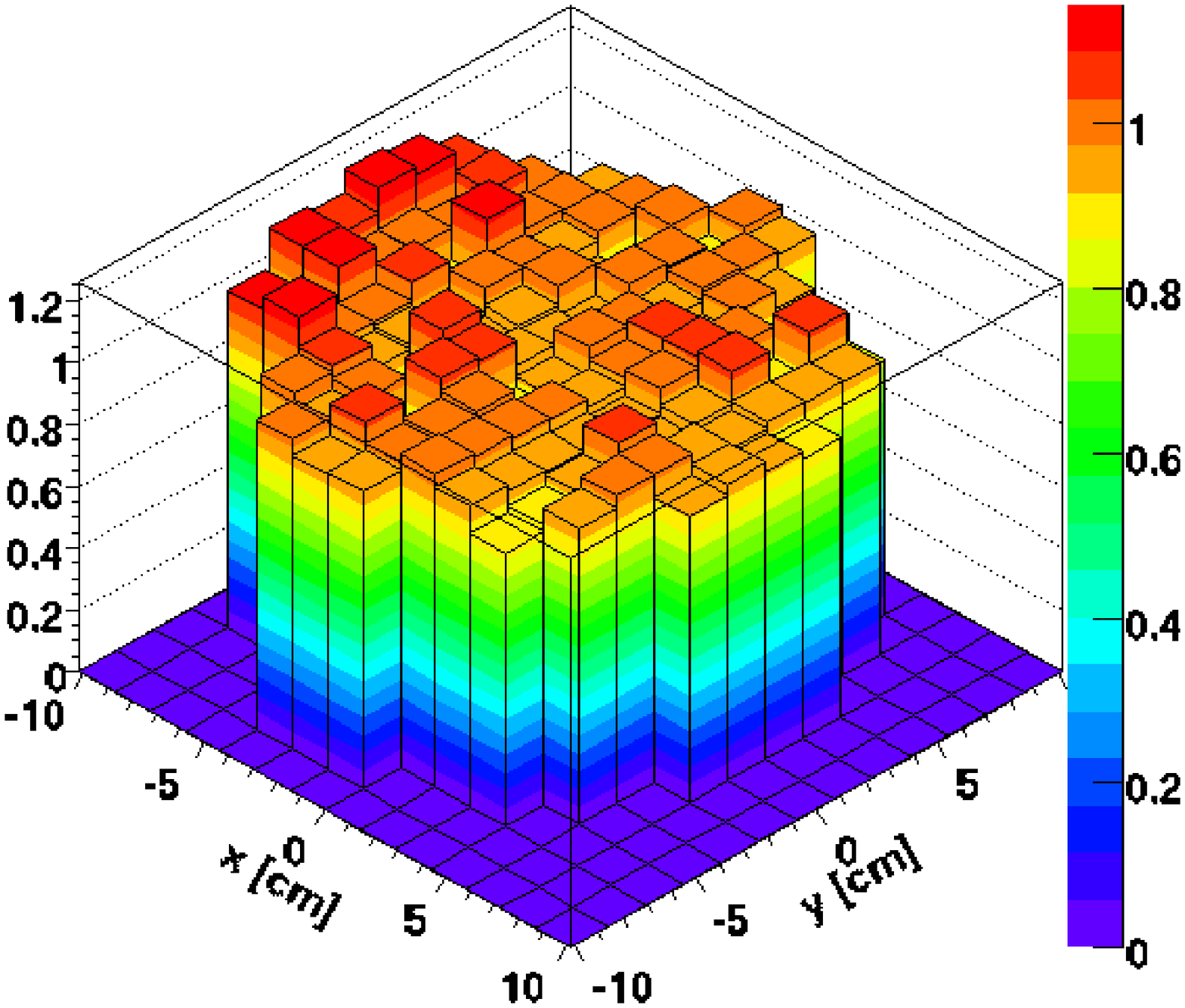}
\includegraphics[height=7.cm,width=7.cm]{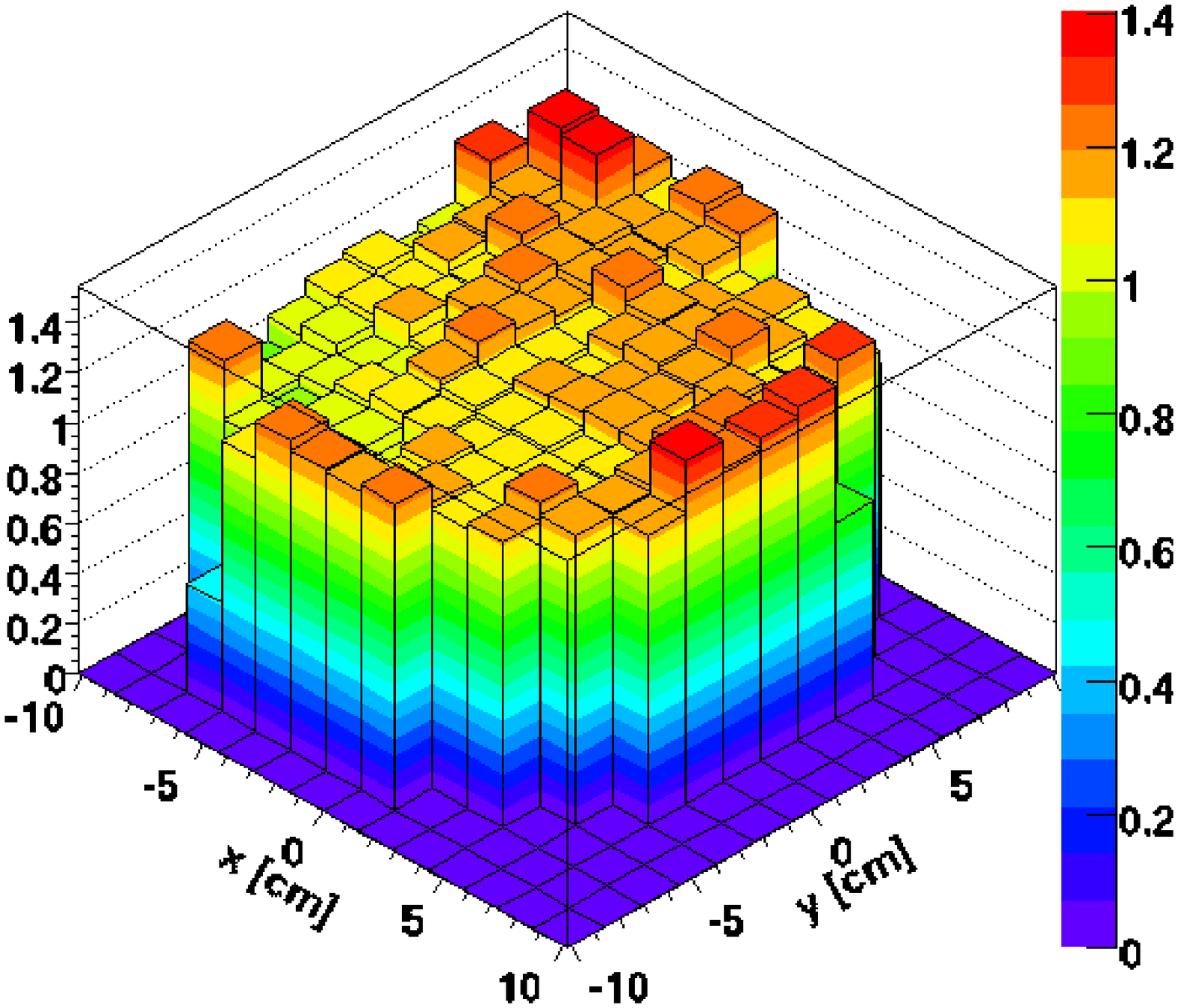}
\caption{Plots of a) The collection efficiency and b) The gain of the PMT 
with a -1 gauss magnetic field along the axis of the PMT.}
\label{axial}
\efi

When an axial magnetic field is induced, this dip is not present, and the 
response remains relatively stable across the photocathode 
(Figure \ref{axial}).  Since a magnetic field in the axial direction 
would not significantly bend the electrons away from the first dynode, 
the changes in the response of the PMT are expected to be much less 
noticeable.  Figure \ref{axial} confirms that even in the
presence of a 1 gauss magnetic field, the amount of decrease in both 
collection efficiency and gain is minimized with an axial magnetic field.

\section{Conclusion}
\label{conclusion}
We have measured the PMT response
to various orientations of external magnetic fields as a function of the 
cathode position impinged by a collimated light source. A clear
dependence on the cathode position in both collection efficiency and
gain of the PMT is observed. Although the effects on
the PMT gain is smaller than that on the PMT collection efficiency, both
effects have been observed. In particular, a pronounced valley in 
the collection efficiency is observed for certain locations of the 
cathode when the magnetic field is transverse to the PMT axis.
An axial magnetic field, however, has been shown to create a relatively 
small change in collection efficiency and gain. This study shows that 
a detailed modelling of the response of a large diameter PMT in the 
presence of an external magnetic field should take into account the 
significant cathode position dependence.

\section*{Acknowledgements}
We gratefully acknowledge the advice and assistance of Professors 
D. Hertzog and K.B. Luk during the course of this measurement.  
This work is supported by the U.S. National Science Foundation.

\end{document}